\documentclass{tlp}

\usepackage{epsfig} 
\usepackage{amssymb} 
\usepackage{latexsym}

\def\onetom{\{1,\dots,m\}}

\newcommand{\pred}{\mathit{pred}} 
\newcommand{\term}{\mathit{term}} 
\newcommand{\float}{\mathit{float}}

\newcommand{\false}{{\mbox {false}}}
\newcommand{\size}{{\mbox {size}}}
\newcommand{\Max}{{\mbox {Max}}} 
\newcommand{\lb}{{\mbox {lb}}} 
\newcommand{\ub}{{\mbox {ub}}} 
\newcommand{\glb}{{\mbox {glb}}} 
\newcommand{\lub}{{\mbox {lub}}} 
\newcommand{\range}{{\mbox {range}}} 
\newcommand{\vars}{{ V}}

\def\ftau{f_{\tau_1 \dots \tau_n\rightarrow \tau}} 
\def\ptau{p_{\tau_1 \dots \tau_n}}

\def\qed{}

\def\X{{\mathcal X}}
\def\F{{\mathcal F}}
\def\K{{\mathcal K}}
\def\T{{\mathcal T}}
\def\U{{\mathcal U}}
\def\V{{\mathcal V}}
\def\P{{\mathcal P}}
\def\G{{\mathcal G}}
\def\N{{\mathcal N}}
\def\R{{\mathcal R}}

\def\la{{\leftarrow}}
\def\ra{{\rightarrow}}
\def\lra{{\longrightarrow}}
\def\CSLD{{\lra_{CSLD}}}
\def\Query{{\mathit{Query}}}
\def\Clause{{\mathit{Clause}}}
\def\A{{\overline A}}
\def\B{{\overline B}}
\def\Atom{{\mathit{Atom}}}
\def\Head{{\mathit{Head}}}

\def\leE{{\le }}

\newtheorem{lemma}{Lemma}[section]
\newtheorem{proposition}{Proposition}[section]
\newtheorem{definition}{Definition}[section]
\newtheorem{theorem}{Theorem}[section]
\newtheorem{example}{Example}[section]
\newtheorem{corollary}{Corollary}[section]

\begin{document} 

\bibliographystyle{tlp} 
 
\title{Typing Constraint Logic Programs}
 
\author[Fran\c{c}ois Fages and Emmanuel Coquery]{
Fran\c{c}ois Fages and Emmanuel Coquery\\
Projet Contraintes, INRIA-Rocquencourt,\\
BP105, 78153 Le Chesnay Cedex, France, \\ 
\email{\{francois.fages,emmanuel.coquery\}@inria.fr}}

\jdate{July 2001}
\pubyear{2001}
\pagerange{\pageref{firstpage}--\pageref{lastpage}}

\maketitle 
\begin{abstract} 
We present a prescriptive type system with parametric
polymorphism and subtyping for constraint logic programs.  The aim of
this type system is to detect programming errors statically. It
introduces a type discipline for constraint logic programs and
modules, while maintaining the capabilities of performing the usual
coercions between constraint domains, and of typing meta-programming
predicates, thanks to the flexibility of subtyping.  The property of
{\em subject reduction} expresses the consistency of a prescriptive type system
w.r.t.\ the execution model: if a program is ``well-typed'', then all
derivations starting from a ``well-typed'' goal are again
``well-typed''.  That property is proved w.r.t. the abstract execution
model of constraint programming which proceeds by accumulation of
constraints only, and w.r.t. an enriched execution model with type constraints
for substitutions.
We describe our implementation of the system for type
checking and type inference. We report our experimental results on
type checking ISO-Prolog, the (constraint) libraries of Sicstus Prolog and other Prolog
programs.
\end{abstract}

\section{Introduction}\label{intro-sec} 

The class CLP($\X$) of Constraint Logic Programming languages
was introduced by Jaffar and Lassez \cite{JL87} as a generalization
of the innovative features introduced by Colmerauer in Prolog II \cite{Colmerauer84,Colmerauer85}:
namely computing in Prolog with other structures than the Herbrand terms,
with inequality constraints and with co-routining.

Inherited from the Prolog tradition,  CLP($\X$) programs are untyped.
Usually the structure of interest $\X$ is however a quite complex
combination of basic structures that may include integer arithmetic,
real arithmetic, booleans, lists, Herbrand terms, infinite terms, etc. 
with implicit coercions between constraint domains like in Prolog IV \cite{Colmerauer96}.
Even the early CLP($\R$) system of \cite{JL87} already combines
Herbrand terms with arithmetic expressions in a non-symmetrical way:
any arithmetic expression may appear under a Herbrand function symbol,
e.g.~in a list, but not the other way around.
The framework of {\em many sorted logic} in \cite{JL87}
is not adequate for representing
the type system underlying such a combination, as it forces Herbrand
function symbols to have a unique type (e.g.~over reals or Herbrand terms),
whereas Herbrand functions can be used polymorphically, e.g.~in
{\tt f(1)} and {\tt f(f(1))}, or the list constructor
in a list of list of numbers {\tt [[3]]}.

The type system of Mycroft-O'Keefe~\cite{MK84,LR91,HT92-new} is an adaptation to logic programming
of the first type system with {\em parametric polymorphism},
that was introduced by Damas-Milner for the functional programming language ML.
In this system, types are first-order terms, type variables inside types,
like $\alpha$ in $list(\alpha)$, express type parameters.
Programs defined over a data structure of type $list(\alpha)$ 
can be used polymorphically over any homogeneous list of elements 
of some type $\alpha$.
Such a type system for Prolog is implemented in the systems G\"{o}del~\cite{HL94}
and Mercury \cite{mercury} for example.
The flexibility of parametric polymorphism is however by far insufficient
to handle properly coercions between constraint domains,
such as e.g.~booleans as natural numbers, or lists as Herbrand terms,
and does not support the meta-programming facilities of logic programming,
with meta-predicates such as {\tt functor(X,F,N), 
call(G)} or {\tt setof(X,G,L)}.

Semantically, a ground type represents a set of expressions.  
Subtyping makes type systems more expressive and flexible in that 
it allows to express inclusions among these sets. 
In this paper we investigate the use of subtyping for expressing
coercions between constraint domains, and for typing meta-programming predicates.
The idea is that by allowing subtype relations like $list(\alpha)\leq term$,
an atom like $functor([X|L],F,N)$ is well-typed
with type declaration $functor: term\times atom\times int\rightarrow \pred$,
although its first argument is a list.
Similarly, we can type $call:\pred\rightarrow \pred$, 
$freeze:term\times \pred\rightarrow \pred$,
$setof:\alpha\times \pred\times list(\alpha)\rightarrow \pred$.
The absence of subtype relation $list(\alpha)\not\leq \pred$,
has for effect to raise a type error
if the {\tt call} predicate is applied to a list.
On the other hand, the subtype relation $\pred\leq term$ makes
coercions possible from goals to terms.

Most type systems with subtyping for logic programming languages 
that have been proposed 
are descriptive type systems, i.e.~their purpose is to describe 
the success set of the program,
they require that a type for a predicate upper approximates its denotation.
On the other hand, in prescriptive type systems,
types are syntactic objects defined by the user to express
the intended use of function and predicate symbols in 
programs. Note that the distinction between descriptive and prescriptive type systems
is orthogonal to the distinction between type checking and type inference
which are possible in both approaches.

There are only few works considering prescriptive type systems for  
logic programs with 
subtyping~\cite{Bei95,DH88,Han92,HT92-new,YFS92,Smo88}.  
In these systems however, subtype relations between parametric type constructors
of different arities,
like $list(\alpha)\leq term$,
are not allowed, 
thus they cannot be used to type metaprogramming predicates
and have not been designed for that purpose.
The system Typical \cite{Mey96} possesses an ad hoc mechanism for typing metapredicates
which makes it quite difficult to use. Our objective is to propose
a simple type system that allows for a uniform treatment of
prescriptive typing issues in constraint logic programs.

In a prescriptive type system, 
the property of {\em subject reduction} expresses the consistency
of the type system w.r.t.\ the execution model:
if a program is ``well-typed'', 
then all derivations starting in a ``well-typed'' goal are again 
``well-typed''. 
This is a well-known result of the polymorphic type system without subtyping 
\cite{MK84,LR91,HT92-new} but when subtypes are added to the picture,
the absence of a fixed data flow in logic programs
makes the obtention of a similar result problematical.
Beierle~\cite{Bei95} shows the existence of 
principal typings with subtype relations between basic types, and provides type inference algorithms, however
Beierle and also 
Hanus~\cite{Han92} do not claim subject reduction for the systems they 
propose. In general types are kept at run-time \cite{Han92,YFS92}
or modes are introduced to restrict the data flow \cite{DH88,SFD00fsttcs,mercury}.

\nocite{FP98}
In this paper, by abstracting from particular structures
as required in the CLP scheme,
we study a prescriptive type system for CLP programs, 
that is independent from any specific constraint domain $\X$.
Section \ref{TS} presents
the type system that includes parametric polymorphism
and subtype relations between type constructors of different arities,
in a quite general type structure of poset with suprema.
We show two subject reductions results.
One is relative to the abstract execution model
of constraint programming, which proceeds only by accumulation of constraints.
The proof of subject reduction holds independently of the computation domain,
under the assumption that the type of predicates
satisfies the definitional genericity principle \cite{LR91}.
The second subject reduction result is relative to
the more concrete execution model of CLP with substitution steps.
We show that for this second form it is necessary
to keep at run-time the typing constraints on variables inside well-typed programs and queries.

Section \ref{TC} describes the type checking algorithm and shows that 
the system of subtype inequalities generated by the type checker are
left-linear and acyclic.
Section \ref{SSI} presents a linear time algorithm for solving left-linear and
acyclic systems of subtype inequalities, and describes the cubic time algorithm
of Pottier \cite{Pottier00ic}~for solving general systems of inequalities,
under the additional assumption that the types form a lattice.
Section \ref{TI} presents type inference algorithms for inferring the types
of variables and predicates in program clauses.

Section~\ref{implementation} describes our implementation 
which is available from \cite{Coquery00tclp}.
The solving of subtype inequalities is done by an interface to the 
Wallace constraint-handling library~\cite{Pottier00wallace}.
In section~\ref{boum} we report our experimental results
on the use of this implementation to type check ISO-Prolog, the libraries of
Sicstus Prolog, including constraint programming libraries, and other Prolog programs.

\section{Typed Constraint Logic Programs}\label{TS}

In this section we describe our type system as a logic for deriving type judgments
about CLP programs.

\subsection{Types}\label{program-symbols-subsec} 

The type system we consider is based on a structure of
partially ordered terms, called poterms,
that we use for representing types with both parametric polymorphism and
subtype polymorphism.
Poterms generalize first-order terms
by the definition of a subsumption order based on function symbols,
that comes in addition to the instantiation preorder based on variables.
Poterms are similar to order-sorted
feature terms or $\psi$-terms \cite{AN86,Smo88,APG97} but we find it more convenient here to adopt
a term syntax (with matching by position) instead of
a record syntax (with matching by name)
for denoting static types.

The set of types $\T$ is the set of terms formed over
a denumerable set $\U$ of {\em type variables}
(also called {\em parameters}), denoted by $\alpha,\ \beta,...$,
a finite set of {\em constructors} $\K$, where with each 
$K\in\mathcal{K}$ an arity 
$m\geq 0$ is associated (by writing $K/m$).
{\em Basic types} are type constructors of arity 0.
We assume that $\K$ contains a basic type $\pred$.
A {\em flat type} is a type of the form  
$K(\alpha_1,\dots,\alpha_m)$, where $K \in \K$ and the $\alpha_i$ are  
distinct parameters. 

The set of type variables in a type $\tau$ is denoted by $V(\tau)$.
The set of ground types $\G$ is the set of types containing no variable.
We write $\tau[\sigma/\alpha]$ to denote the type obtained by replacing
all the occurrences of $\alpha$ by $\sigma$ in $\tau$.
We write $\tau[\sigma]$ to denote that the type $\tau$ 
strictly contains the type $\sigma$ as a subexpression. 
The size of a type $\tau$, defined as the number of occurrences of constructors and parameters in $\tau$,
is denoted by $\size(\tau)$.
 
We now qualify what kind of subtyping we allow. Intuitively, when a
type $\sigma$ is a subtype of a type $\tau$, this means that each term
in $\sigma$ is also a term in $\tau$. The subtyping relation $\leq$ is
designed to have certain nice algebraic properties, stated in
propositions below.
We assume an 
order $\leq$ on type constructors 
such that: $K/m\leq K'/m'$ implies $m\geq m'$,
and for each $K\in\K$ the set  
$\{K' \mid K\leq K'\}$ has a maximum. 
Moreover, we assume that with each pair 
$K/m\leq K'/m'$, an injective mapping 
$\iota_{K,K'}: \{1,\dots,m'\} \rightarrow \onetom$ 
is associated such that $\iota_{K,K''}=\iota_{K,K'}\circ\iota_{K',K''}$
whenever $K\le K'\le K''$.

These assumptions mean that as we move up in the hierarchy of type constructors,
their arity decreases, and the hierarchy needs not be a lattice but a poset with suprema.

The order on type constructors is extended to a structural {\em covariant subtyping order} on
types, denoted also by $\leq$, defined as the least relation satisfying the 
following rules:

\begin{center}
\begin{tabular}{lll}
{\em (Par)} & 
$\alpha\leq \alpha$ & 
$\alpha$ is a parameter\\[2ex]
{\em (Constr)} &
\Large
$\frac%
  {\tau_{\iota(1)}\leq\tau'_1\ \dots \ \tau_{\iota(m')}\leq\tau'_{m'}}%
  {K(\tau_1,\dots,\tau_m)\leq K'(\tau'_1,\dots,\tau'_{m'})}\quad$ &
$K\leq K'$, $\iota = \iota_{K, K'}$.
\end{tabular}
\end{center}

Contravariant type constructors could be defined with a subtyping rule
similar to rule {\em Constr} but with the ordering relation reversed
for some arguments, like e.g.~$\tau_{\iota(i)}\geq\tau'_i$
in the premise of the rule for some argument $\tau'_i$.
Such contravariant type constructors are not considered in this paper.

Therefore, if $int\leq float$ then we have
$list(int)\leq list(float)$, $list(float)\not\leq list(int)$,
and also $list(float)\not\leq list(\alpha)$ as the subtyping order does not include 
the instantiation pre-order.
Intuitively, a ground type represents a set of expressions,
and the subtyping order between ground types corresponds to set inclusion.
Parametric types do not directly support this interpretation as it
would identify all parameters.

\begin{center}
\begin{figure}\label{iso-prolog}
\begin{center}
\epsfig{file=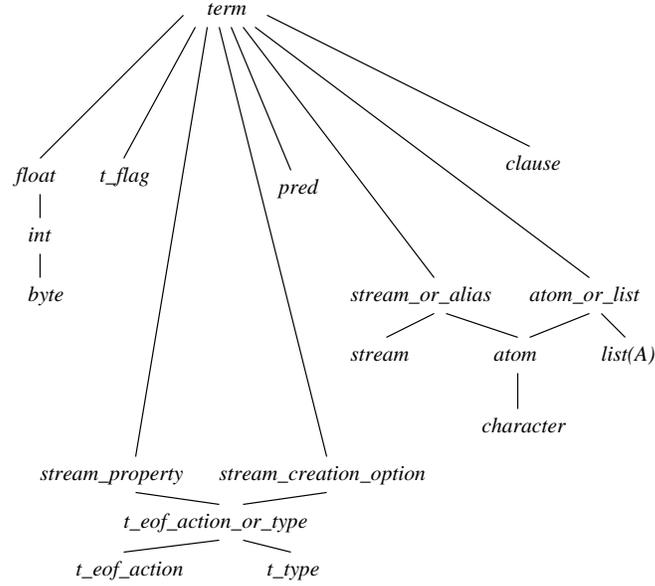}
\end{center}
\caption{Part of the type structure for ISO-Prolog.}
\end{figure}
\end{center}

The type structure given in figure~\ref{iso-prolog}
represents a part of the types used for type checking ISO-Prolog.
The omitted types are the subtypes of {\em atom} associated to all types,
and other types for special values or options.
The type $list(\alpha)$ is the only parametric type used for ISO-Prolog.
Other parametric types are used for typing Prolog libraries such as 
$arrays(\alpha)$, $assoc(\alpha,\beta)$, $heaps(\alpha,\beta)$,
$ordsets(\alpha)$, $queues(\alpha)$, etc.

A {\em type substitution} $\Theta$ is an idempotent mapping 
from parameters to types that is the identity almost everywhere. 
Applications of type substitutions are defined 
in the obvious way.

\begin{proposition}\label{inst-closed-prop}
If $\sigma\leq\tau$ then $\sigma\Theta\leq\tau\Theta$ for any type substitution $\Theta$.
\end{proposition}
\begin{proof}
By structural induction on $\tau$.
\qed\end{proof}

\begin{proposition}\label{size-prop}
If $\sigma\leq\tau$ then $\size(\sigma)\geq \size(\tau)$ .
\end{proposition}
\begin{proof}
By structural induction on $\tau$.
\qed\end{proof}

Our assumption that for each $K\in\K$, the set  
$\{K' \mid K\leq K'\}$ has a maximum, 
together with the arity decreasing assumption,
entail the existence of 
a maximum supertype for any type:

\begin{proposition}\label{max-exists} 
For each type $\tau$, the set
$\{\sigma \mid \tau\leq\sigma\}$ has a maximum, which is denoted by 
$\Max(\tau)$.
\end{proposition}
\begin{proof}
By structural induction on $\tau$.
\qed\end{proof}

This means that every $\le$-connected
component of types has a root.
For example, a structure like $a\le b,\ c\le b, c\le d$
violates the hypothesis if $b$ and $d$ have no common supertype
serving as a root for the connected component.
On the other hand that assumption does not assume, nor it is implied by,
the existence of a least upper bound to types having a upper bound
(sup-quasi-lattice hypothesis in \cite{Smo89}).

\begin{proposition}\label{max-prop} 
For 
all types $\tau$ and $\sigma$, $\Max(\tau[\sigma/\alpha])=\Max(\tau)[\Max(\sigma)/\alpha]$.
\end{proposition}
\begin{proof}
By structural induction on $\tau$.
\qed\end{proof}

Note that the possibility of ``forgetting'' type parameters in
subtype relations, as in $list(\alpha)\le term$, may provide solutions
to inequalities of the form $list(\alpha)\le \alpha$, e.g.~$\alpha=term$.
However, we have:

\begin{proposition}\label{propfini}\label{corfini}
An inequality of the form $\alpha\le \tau[\alpha]$ has no solution.
An inequality of the form $\tau[\alpha]\le \alpha$ has no solution if $\alpha\in \vars(\Max(\tau))$.
\end{proposition}
\begin{proof}
For any type $\sigma$, we have $\size(\sigma)<\size(\tau[\sigma])$,
hence by Prop~\ref{size-prop}, $\sigma\not\le\tau[\sigma]$,
that is $\alpha\le \tau[\alpha]$ has no solution.

For the second proposition, we prove its contrapositive.
Suppose $\tau[\alpha]\le \alpha$ has a solution, say $\tau[\sigma/\alpha]\le \sigma$.
By definition of a maximum and Prop.~\ref{max-exists}, we have $\Max(\sigma)=\Max(\tau[\sigma/\alpha])$.
Hence by Prop.~\ref{max-prop}, $\Max(\sigma)=\Max(\tau)[\Max(\sigma)/\alpha]$.
By the rules of subtyping we have
$\alpha\neq\Max(\tau)$. Therefore $\alpha\not\in\vars(\Max(\tau))$, 
since otherwise $\Max(\sigma)=\Max(\tau)[\Max(\sigma)/\alpha]$ would contain 
$\Max(\sigma)$ as a strict subexpression which is impossible.
\qed\end{proof}

\subsection{Well-typed programs}\label{typed-language-subsec}

CLP programs are built over a denumerable set $\V$ of {\em variables},
a finite set $\F$ 
of {\em function} symbols, given with their arity
(constants are functions of arity 0),
and a finite set $\P$ of {\em program predicate} and {\em constraint predicate}
symbols given with their arity, containing the equality constraint $=$.
A query $Q$ is a finite sequence of constraints and atoms.
A program clause is an expression noted $A\leftarrow Q$ where $A$
is an atom formed with a program predicate and $Q$ a query.

A {\em type scheme} is an expression of the form
$\forall\overline\alpha\tau_1,\dots,\tau_n\ra\tau$, 
where $\overline\alpha$ is the set of parameters
in types $\tau_1,...,\tau_n,\tau$.
We assume that each function symbol $f \in \F$,
has a {\em declared type scheme} of the form
$\forall\overline\alpha\tau_1,\dots,\tau_n\ra\tau$, 
where $n$ is the arity of $f$, and $\tau$ is a flat type.
Similarly, we assume that each predicate symbol $p\in\P$
has a declared type scheme of the form
$\forall\overline\alpha\tau_1,\dots,\tau_n\ra\pred$
where $n$ is the arity of $p$.
The declared type of the equality constraint symbol is
$\forall u\ u,u\ra\pred$.
For notational convenience,
the quantifiers in type schemes and the resulting type $\pred$
of predicates will be omitted in type declarations, 
the declared type schemes will be indicated by writing 
$\ftau$ and $\ptau$, assuming a fresh renaming of the parameters in
$\tau_1,\dots,\tau_n,\tau$ for each occurrence of $f$ or $p$.

Throughout this paper, 
we assume that $\K$, $\F$, and 
$\P$ are fixed by means of declarations in a 
{\em typed program}, where the syntactical
details are insignificant for our results.

A {\em variable typing} 
is a mapping from a finite subset of $\V$ to $\T$,
written as $\{x_1:\tau_1,\dots,x_n:\tau_n\}$. 
The type system defines well-typed terms, atoms and clauses
relatively to a variable typing $U$. 
The typing rules are given in Table \ref{rules-tab}.
The rules basically consist of the rules of Mycroft and O'Keefe plus a subtyping rule.
Note that for the sake of simplicity
constraints are not distinguished from other atoms in this system.

\begin{table}[htb]
\begin{center}
\begin{tabular}{lll}
{\em (Sub)} &
\Large
$\frac{U\vdash t:\tau\ \tau\leq\tau'}{U\vdash t:\tau'}$\\[1ex]
&\\
{\em (Var)} &
$\{x:\tau,\dots\}\vdash x:\tau$\\[2ex]
&\\
{\em (Func)} &
\Large
$\frac%
  {U\vdash t_1:\tau_1\Theta \ \dots \ U\vdash t_n:\tau_n\Theta}
  {U\vdash\ftau(t_1,\dots,t_n):\tau\Theta}$ &
$\Theta$ is a type substitution\\[2ex]
&\\
{\em (Atom)} &
\Large
$\frac%
  {U\vdash t_1:\tau_1\Theta \ \dots \ U\vdash t_n:\tau_n\Theta}
  {U\vdash\ptau(t_1,\dots,t_n) \mathit{Atom}}$ &
$\Theta$ is a type substitution\\[2ex]
&\\
{\em (Head)} &
\Large
$\frac%
  {U\vdash t_1:\tau_1\Theta \ \dots \ U\vdash t_n:\tau_n\Theta}
  {U\vdash\ptau(t_1,\dots,t_n) \mathit{Head}}$ &
$\Theta$ is a renaming substitution\\[2ex]
&\\
{\em (Query)} &
\Large
$\frac%
  {U\vdash A_1\ \mathit{Atom}\ \dots \  U\vdash A_n\ \mathit{Atom}}%
  {U\vdash A_1,\dots,A_n\ \mathit{Query}}$ \\[2ex]
&\\
{\em (Clause)} & 
\Large
$\frac%
  {U\vdash Q\ Query \quad U\vdash A\ \mathit{Head}}%
  {U \vdash A \leftarrow Q\ \mathit{Clause}}$\\[2ex]
&\\
\end{tabular}
\end{center}
\caption{The type system.\label{rules-tab}}
\end{table}

An object, say a term $t$, is {\em well-typed} if there exist
some variable typing $U$ and some type $\tau$ such that $U\vdash t:\tau$.
Otherwise the term is {\em ill-typed}
(and likewise for atoms, etc.). 
A program is well-typed if all its clauses are well-typed.

The distinction between rules {\em Head} and {\em Atom}
expresses the usual {\em definitional genericity} principle \cite{LR91}
which states that the type of a defining occurrence of a predicate
(i.e.~at the left of ``$\leftarrow$'' in a clause)
must be equivalent up-to renaming to the assigned type of the predicate.
The rule {\em Head} used for deriving the type of the head of the clause
is thus not allowed to use substitutions other than
variable renamings in the declared type of the predicate.
For example, the predicate $member$ can be typed polymorphically,
i.e.~$member: \alpha\times list(\alpha)\ra \pred$,
if its definition does not contain special facts like $member(1,[1])$,
that would force its type to be 
$member: int\times list(int)\ra \pred$,
for satisfying the definitional-genericity condition.

The following proposition shows that if an expression other than a clause or a head
is well-typed in a variable typing $U$, it remains well-typed
in any instance $U\Theta$.

\begin{proposition}\label{instance}
For any variable typing $U$, any type judgement $R$ other than a Head or a Clause, 
and any type subtitution $\Theta$, if $U\vdash R$ then $U\Theta\vdash R\Theta$.
\end{proposition}
\begin{proof}
By induction on the height of the derivation tree for $U\vdash R$.
\qed\end{proof}

\subsection{Subject reduction w.r.t. CSLD resolution}

Subject reduction is the property that evaluation rules transform a
 well-typed expression into another well-typed expression. The
 evaluation rule for constraint logic programming is CSLD-resolution.
To recall this evaluation rule, it is convenient to distinguish in a query $Q$, the constraint
part $c$ (where the sequence denotes the conjunction) 
from the other sequence of atoms $\A$. We use the notation $Q=c|\A$ to make this distinction.
Given a constraint domain $\X$ which fixes the interpretation of constraints,
a query $c'|\B$ is a {\em CSLD-resolvent} of a query $c|\A$ and a (renamed apart) program clause
$p(t_1,...,t_n)\leftarrow d|\A$, if

$\A= A_1,\dots,A_{k-1},p(t'_1,\ldots,t'_n),A_{k+1},\ldots,A_m$,  

$\B=A_1,\dots,A_{k-1},\A,A_{k+1},\dots,A_m$,  

and the constraint $c'=(c\wedge d\wedge t_1=t'_1\wedge\ldots\wedge t_n=t'_n)$ is $\X$-satisfiable.

\begin{theorem}[Subject Reduction for CSLD resolution]\label{subject-reduction} 
Let $P$ be a well-typed CLP($\X$) program,
and $Q$ be a well-typed query, i.e.~$U\vdash Q\ Query$ for some variable typing $U$.
If $Q'$ is a CSLD-resolvent of $Q$, then there exists a variable typing $U'$
such that $U'\vdash Q'\ Query$.
\end{theorem}
\begin{proof}
Let us assume without loss of generality that
$Q=c|p(s),\A$, and that $Q'$ is a CSLD-resolvent of $Q$ with the program  clause $p(t)\la d|\B$.

Thus $Q'=c,d,s=t|\A,\B$.

As $Q$ is well-typed, we have $U\vdash c|p(s),\A\ \Query$.
And as the program is well typed, there exists a variable typing $U''$,
renamed apart from $U$, such that $U''\vdash p(t)\la d|\B\ \Clause$.

Let $p:\tau\ra\pred$ be the type declaration of predicate $p$. Since $U \vdash p(s)\ \Atom$,
we have $U\vdash s:\tau\Theta$ for some substitution $\Theta$. 

Now let $U'=U\cup U''\Theta$.
By proposition~\ref{instance}, we have 
$U''\Theta\vdash d|\B,\ \Query$,
thus $U'\vdash c,d|\A,\B,\ \Query$,
What remains to be shown is $U'\vdash s=t\ \Atom$.

Since $U''\vdash p(t)\ \Head$,
we have $U''\vdash t:\tau$.
Hence by proposition~\ref{instance}, $U''\Theta\vdash t:\tau\Theta$.
Therefore we have $U\vdash s:\tau\Theta$ and $U''\Theta\vdash t:\tau\Theta$,
from which we conclude $U'\vdash s=t\ \Atom$.
\qed\end{proof}

It is worth noting that the previous result
would not hold without
the definitional genericity condition (expressed in rule {\em Head}).
For example with two constants $a:\tau_a$ and $b:\tau_b$, and
one predicate $p:\alpha\ra \pred$ defined by the non definitional generic clause $p(a)$,
we have that the query $p(b)$ is well typed, but
$b=a$ is a resolvent that is ill-typed if $\tau_a$ and $\tau_b$ have no upper bound.

\subsection{Subject reduction w.r.t. substitutions}

The CSLD reductions, noted $\CSLD$, 
are in fact an abstraction of the operational reductions that may perform also
substitution steps, noted $\lra_\sigma$,
instead of keeping equality constraints.
As in the CLP scheme constraints are handled modulo logical equivalence \cite{JL87},
it is clear that the diagram of both reductions commutes~:
\begin{center}
\setlength{\unitlength}{3947sp}%
\begingroup\makeatletter\ifx\SetFigFont\undefined%
\gdef\SetFigFont#1#2#3#4#5{%
  \reset@font\fontsize{#1}{#2pt}%
  \fontfamily{#3}\fontseries{#4}\fontshape{#5}%
  \selectfont}%
\fi\endgroup%
\begin{picture}(3300,2265)(901,-1861)
\put(901,239){\makebox(0,0)[lb]{\smash{\SetFigFont{12}{14.4}{\rmdefault}{\mddefault}{\updefault}
$Q_1$
}}}
\put(2101,-61){\makebox(0,0)[lb]{\smash{\SetFigFont{12}{14.4}{\rmdefault}{\mddefault}{\updefault}
$\downarrow_\sigma$
}}}
\put(901,-61){\makebox(0,0)[lb]{\smash{\SetFigFont{12}{14.4}{\rmdefault}{\mddefault}{\updefault}
$\downarrow_\sigma$
}}}
\put(1201,239){\makebox(0,0)[lb]{\smash{\SetFigFont{12}{14.4}{\rmdefault}{\mddefault}{\updefault}
$\CSLD$
}}}
\put(1201,-361){\makebox(0,0)[lb]{\smash{\SetFigFont{12}{14.4}{\rmdefault}{\mddefault}{\updefault}
$\CSLD$
}}}
\put(4201,239){\makebox(0,0)[lb]{\smash{\SetFigFont{12}{14.4}{\rmdefault}{\mddefault}{\updefault}
$Q_n$
}}}
\put(4201,-61){\makebox(0,0)[lb]{\smash{\SetFigFont{12}{14.4}{\rmdefault}{\mddefault}{\updefault}
$\downarrow_\sigma$
}}}
\put(2101,239){\makebox(0,0)[lb]{\smash{\SetFigFont{12}{14.4}{\rmdefault}{\mddefault}{\updefault}
$Q_2$
}}}
\put(2101,-661){\makebox(0,0)[lb]{\smash{\SetFigFont{12}{14.4}{\rmdefault}{\mddefault}{\updefault}
$\downarrow_\sigma$
}}}
\put(4201,-661){\makebox(0,0)[lb]{\smash{\SetFigFont{12}{14.4}{\rmdefault}{\mddefault}{\updefault}
$\downarrow_\sigma$
}}}
\put(4201,-1261){\makebox(0,0)[lb]{\smash{\SetFigFont{12}{14.4}{\rmdefault}{\mddefault}{\updefault}
...
}}}
\put(4201,-1561){\makebox(0,0)[lb]{\smash{\SetFigFont{12}{14.4}{\rmdefault}{\mddefault}{\updefault}
$\downarrow_\sigma$
}}}
\put(4201,-1861){\makebox(0,0)[lb]{\smash{\SetFigFont{12}{14.4}{\rmdefault}{\mddefault}{\updefault}
$Q$
}}}
\put(2401,239){\makebox(0,0)[lb]{\smash{\SetFigFont{12}{14.4}{\rmdefault}{\mddefault}{\updefault}
$\CSLD$
}}}
\put(2401,-961){\makebox(0,0)[lb]{\smash{\SetFigFont{12}{14.4}{\rmdefault}{\mddefault}{\updefault}
$\CSLD$
}}}
\put(3201,239){\makebox(0,0)[lb]{\smash{\SetFigFont{12}{14.4}{\rmdefault}{\mddefault}{\updefault}
...
}}}
\put(3401,239){\makebox(0,0)[lb]{\smash{\SetFigFont{12}{14.4}{\rmdefault}{\mddefault}{\updefault}
$\CSLD$
}}}
\put(3201,-361){\makebox(0,0)[lb]{\smash{\SetFigFont{12}{14.4}{\rmdefault}{\mddefault}{\updefault}
...
}}}
\put(3201,-1261){\makebox(0,0)[lb]{\smash{\SetFigFont{12}{14.4}{\rmdefault}{\mddefault}{\updefault}
...
}}}
\end{picture}

\end{center}

However the previous subject reduction result expresses the
consistency of types w.r.t. horizontal reduction steps only,
that is w.r.t.~the abstract execution model which accumulates
constraints, but may not hold for more concrete operations of constraint solving
and substitutions.
For example, with the subtype relations $int\leq term$, $\pred\leq term$, the 
type declarations $=:\alpha\times\alpha\ra \pred$, $p:int\ra \pred$,
and the program $p(X)$,
the query $Y=true, p(Y)$ is well typed with $Y:int$, and succeeds with $Y=true$,
although the query obtained by substitution, $p(true)$, is ill-typed.
In order to establish subject reduction for substitution steps,
and be consistent with the semantical equivalence of programs,
one needs to consider a typed execution model with 
type constraints on variables checked at runtime.
In the example, the type constraint $Y:int$ 
with the constraint $Y=true$ is unsatisfiable,
the query can thus be rejected at compile-time by checking the satisfiability of its typed constraints.

\begin{definition}
Given a constraint system over some domain $\X$,
a {\em typed constraint system} over $\X\cup 2^\X$ is defined by adding
type constraints, i.e.~expressions of the form $t:\tau$
where $t$ is a term and $\tau$ a type.
Basic types are interpreted by distinguished subsets of $\X$
and type constructors by mappings between subsets of $\X$ 
satisfying the subtyping relation $\leq$ and the type declarations
for function and predicate symbols.
A type constraint $t:\tau$ is satisfiable if there exists
a valuation $\rho$ of the variables in $t$ and the free parameters in $\tau$
such that $t\rho\in\tau\rho$.
A typed constraint system composed  of type constraints and constraints over $\X$
is satisfiable if there exists a valuation which satisfies all constraints
of the system.
\end{definition}


\begin{lemma}\label{type-constraint-lemma}
In a typed constraint system, $X:\tau\wedge X=t$ entails $t:\tau$.
\end{lemma}
\begin{proof}
For any valuation $\rho$, if $X\rho\in\tau\rho$ and $X\rho=t\rho$
then $t\rho\in\tau\rho$.
\qed\end{proof}

\begin{definition}
The TCLP clause (resp. query)  associated to a well-typed program (resp. query)
in a typed environment $U$ is the clause (resp. query) augmented
with the type constraints in $U$.
\end{definition}

\begin{theorem}[Subject Reduction for substitutions]\label{subject-reduction2} 
Let $P$ be a TCLP program associated to well-typed CLP($\X$) program,
and $Q$ be a TCLP query, we have $U\vdash Q\ Query$ for some variable typing $U$.
If $Q'$ is a CSLD-resolvent of $Q$, then the variable typing $U'$ associated to
the type constraints in $Q'$ gives $U'\vdash Q'\ Query$.
Furthermore if $Q'$ contains an equality constraint $X=t$ then $U'\vdash Q'[t/X]\ Query$.
\end{theorem}
\begin{proof}
Subject reduction for CSLD resolution follows from theorem \ref{subject-reduction}
as TCLP programs are just a special case of well-typed CLP programs. Furthermore one easily
checks that the type constraints in $Q'$, that come from the type constraints
in $Q$ and from the resolving TCLP clause, give exactly the 
type environment $U'$ constructed in the proof of the previous theorem,
thus $U'\vdash Q' Query$.

Now let $X=t$ be a constraint in a resolvent $Q'$. Let $X:\tau\in U'$. 
We have $X:\tau$ in the constraint part of $Q'$
which together with $X=t$ entails $t:\tau$ by lemma \ref{type-constraint-lemma}.
Therefore it is immediate from the typing rules that by replacing $X$ by $t$
in the derivation of $U'\vdash Q'\ Query$,
and by completing the derivation with the derivation of $t:\tau$ instead of $X:\tau$,
we get a derivation of $U'\vdash Q'[t/X]\ Query$. 
\qed\end{proof}

The effect of type constraints in TCLP programs is to prevent
the derivation of ill-typed queries by substitution steps.
In addition, queries such as $X:int, X=true, p(X)$
can be rejected at compile-time because of the unsatisfiability of their constraints.
Similarly TCLP program clauses having unsatisfiable typed constraints can be
rejected at compile-time.

Note that in \cite{SFD00fsttcs} another result of subject reduction 
for substitutions is shown without the addition of type constraints 
but in a very restricted context of moded logic programs.

\section{Type checking}\label{TC}

The system described by the rules of Table \ref{rules-tab} is non-deterministic, 
since the rule {\em Sub} can be used anywhere in a typing derivation.
One can obtain a deterministic type checker, directed by the
syntax of the typed program, simply by
replacing the rule {\em Sub} by variants of the rules {\em Func}, {\em Atom} and {\em Head}
with the subtype relation in their premises.
This leads to the following type system in table \ref{rules2-tab}. 

\begin{table}[htb]
\begin{center}
\begin{tabular}{lll}
{\em (Var)} &
$\{x:\tau,\dots\}\vdash x:\tau$\\[2ex]
&\\
{\em (Func')} &
\Large
$\frac%
  {U\vdash t_1:\sigma_1\ \sigma_1\leq\tau_1\Theta\ \ldots\ U\vdash t_n:\sigma_n\ \sigma_n\leq\tau_n\Theta}%
  {U\vdash\ftau(t_1,\dots,t_n):\tau\Theta}$ &
$\Theta$ is a type substitution\\[2ex]
&\\
{\em (Atom')} &
\Large
$\frac%
  {U\vdash t_1:\sigma_1\ \sigma_1\leq\tau_1\Theta\ \ldots\ U\vdash t_n:\sigma_n\ \sigma_n\leq\tau_n\Theta}%
  {U\vdash\ptau(t_1,\dots,t_n) \mathit{Atom}}$ &
$\Theta$ is a type substitution\\[2ex]
&\\
{\em (Head')} &
\Large
$\frac%
  {U\vdash t_1:\sigma_1\ \sigma_1\leq\tau_1\Theta\ \ldots\ U\vdash t_n:\sigma_n\ \sigma_n\leq\tau_n\Theta}%
  {U\vdash\ptau(t_1,\dots,t_n) \mathit{Head}}$&
$\Theta$ is a renaming substitution\\[2ex]
&\\
{\em (Query)} &
\Large
$\frac%
  {U\vdash A_1\ \mathit{Atom}\ \dots \  U\vdash A_n\ \mathit{Atom}}%
  {U\vdash A_1,\dots,A_n\ \mathit{Query}}$ \\[2ex]
&\\
{\em (Clause)} & 
\Large
$\frac%
  {U\vdash Q\ Query \quad U\vdash A\ \mathit{Head}}%
  {U \vdash A \leftarrow Q\ \mathit{Clause}}$
\end{tabular}
\end{center}
\caption{The type system in second form.\label{rules2-tab}}
\end{table}
\begin{proposition}\label{typeequiv}
A program is well typed in the original system if and only if it is well typed in the new one.
\end{proposition}

\begin{proof}
Clearly, if a program is typable in the new system, it is
typable in the original one: one has just to replace
every occurrence of the ({\it Func'}) and ({\it Atom'}) rules respectively with
the following derivations:
\begin{center}
\begin{tabular}{lcl}
 & $
(Func)\frac{\displaystyle(Sub)\frac{\displaystyle\  U \vdash t_{1} : \tau_{1}\:\:\:\tau_{1} \leq \tau_{1}'\Theta}{\displaystyle\  U \vdash t_{1} : \tau_{1}'\Theta} \: \cdot \:\cdot \:\cdot \: \displaystyle(Sub)\frac{\displaystyle\  U \vdash t_{n} : \tau_{n}\:\:\:\tau_{n} \leq \tau_{n}'\Theta}{\displaystyle\  U \vdash t_{n} : \tau_{n}'\Theta}}
{\displaystyle  U \vdash f_{\tau_{1}' \cdot \cdot \times \tau_{n}' \rightarrow \tau'}(t_{1}, \cdot \cdot \cdot ,t_{n}) : \tau'\Theta} $ & \\
 & \\
 & $
(Atom)\frac{\displaystyle(Sub)\frac{\displaystyle\  U \vdash t_{1} : \tau_{1}\:\:\:\tau_{1} \leq \tau_{1}'\Theta}{\displaystyle\  U \vdash t_{1} : \tau_{1}'\Theta} \: \cdot \:\cdot \:\cdot \:
\displaystyle(Sub)\frac{\displaystyle\  U \vdash t_{n} : \tau_{n}\:\:\:\tau_{n} \leq \tau_{n}'\Theta}{\displaystyle\  U \vdash t_{n} : \tau_{n}'\Theta}}
{\displaystyle  U \vdash p_{\tau_{1}' \times \cdot \cdot \times \tau_{n}'}(t_{1}, \cdot \cdot \cdot ,t_{n})\: Atom} $ & \\
 & \\
 & $
(Head)\frac{\displaystyle(Sub)\frac{\displaystyle\  U \vdash t_{1} : \tau_{1}\:\:\:\tau_{1} \leq \tau_{1}'\Theta}{\displaystyle\  U \vdash t_{1} : \tau_{1}'\Theta} \: \cdot \:\cdot \:\cdot \:
\displaystyle(Sub)\frac{\displaystyle\  U \vdash t_{n} : \tau_{n}\:\:\:\tau_{n} \leq \tau_{n}'\Theta}{\displaystyle\  U \vdash t_{n} : \tau_{n}'\Theta}}
{\displaystyle  U \vdash p_{\tau_{1}' \times \cdot \cdot \times \tau_{n}'}(t_{1}, \cdot \cdot \cdot ,t_{n})\: Atom} $ & \\
 & \\
\end{tabular}
\end{center}

Conversely, if a program is typable in the original system, it
is typable in the second one, noted here $\vdash_2$. The proof is
by induction on the typing derivation in the original system. The rules {\it (Var)}, {\it (Query)} and {\it (Clause)} remain the same. The rule {\it (Atom)} and {\it (Head)} are similar to  rule ({\it Func}). We thus show the property for any term $t$~:
if $U\vdash t : \tau$ in the first system, then $U \vdash_2 t : \tau'$ in the second system with $\tau'\leq\tau$.
 
Let us consider the three possible cases, either the proof terminates by the application of the ({\it Var}) rule, by the application of the ({\it Func}) rule or by application of the ({\it Sub}) rule.
 
The first case is trivial as the rule ({\it Var}) is the same in both systems.
 
In the second case, according to the ({\it Func}) rule,
$U\vdash t_{1} : \tau_{1}\Theta \: \cdot \cdot \cdot U\vdash t_{n} : \tau_{n}\Theta$.
Then, by the induction hypothesis, the terms $ t_{1} \: \cdot \cdot \cdot \: t_{n}$ are also type checked to
$U\vdash_2 t_{1} : \tau'_{1} \: \cdot \cdot \cdot U\vdash_2 t_{n} : \tau'_{n}$ by the second system, with $\tau'_{i}\leq\tau_{i}\Theta, i=1..n$.
By applying the ({\it Func'}) rule with $\tau_{i}'=\sigma_{i}, i=1..n$,
we get  $U\vdash_2 f(t_{1}, \cdot \cdot \cdot ,t_{n}):\tau\Theta$.
 
In the third case, according to the {\it (Sub)} rule, $U\vdash t : \tau$ and $\tau\leq\tau'$ allows us to deduce $U\vdash t : \tau'$. By induction hypothesis, t is type checked to $U\vdash_2 t : \sigma$ in the second system, where $\sigma\leq\tau$. Since $\tau\leq\tau'$, we have $\sigma\leq\tau'$. So t is type checked to $U\vdash_2 t : \sigma$, where $\sigma\leq\tau'$.                        
\qed\end{proof}

The construction of the substitution $\Theta$ needed
in rules {\em(Func'), (Atom')} and {\em (Head')} for type checking, 
can be done by solving the system of subtype inequalities collected
along the derivation of a type judgement.
The parameters in the type environment (i.e.~the parameters 
in the types of variables)
are however not under the scope of these substitutions, as they
act only on the parameters
of the (renamed apart) type declarations for function and predicate symbols.
We are thus looking for type substitutions with a restricted domain.
For the sake of simplicity however, instead of dealing formally
with the domain of type substitutions,
we shall simply assume that the parameters in the type of variables are replaced 
by new constants for checking the satisfiability of subtype inequalities,
and avoid unsound instantiations.

Now let $\Sigma$ be
the collection of subtype inequalities $\le$ imposed on types
by rules {\em (Func')} {\em (Atom')} and {\em (Head')} in a derivation.
Let us define the size of a system of inequalities
as the number of symbols.
The size of the system $\Sigma$ of inequalities
associated to a typed program
is $O(nvd)$ where 
$v$ is the size of the type declarations for variables in the program,
$n$ is the size of the program,
and $d$ is the size of the type declarations for function and predicate symbols.

As the type system is deterministic we have:

\begin{proposition}
A well-formed program is typable if and only 
the system of inequalities
collected along its derivation is satisfiable.
\end{proposition}

It is worth noting that the system of inequalities $\Sigma$ collected
in this way for type checking have in fact a very particular form.

\begin{definition}
A system $\Sigma$ of inequalities
is {\em left-linear} if any type variable has at most
one occurrence at the left of $\le$ in the system.
$\Sigma$ is {\em acyclic} if there exists a ranking function on type variables $r:{\U}\rightarrow \N$
such that if $\sigma\le\tau\in\Sigma$, $\alpha\in V(\sigma)$ and $\beta\in V(\tau)$
then $r(\alpha)< r(\beta)$.
\end{definition}

\begin{proposition}\label{acleli}
The system of inequalities generated by the type checking
algorithm is acyclic and left-linear.
\end{proposition}
\begin{proof}
As the type variables in the types of CLP variables have been 
renamed into constants, the only type variables occurring in $\Sigma$
are introduced by rules (Func') (Atom') and (Head'),
and come from (renamed apart) type declarations of function
and predicate symbols.
We can thus associate to each type variable $\alpha$ a rank $h(\alpha)$
defined as the height of its introduction node in the derivation tree
(i.e.~the maximal distance from the node to its leaves).
Now a rule (Func'), (Atom') or (Head') at height $h$ posts inequalities of the form
$\sigma\le\tau$, where the rank of the variables in $\tau$
is $h$, and the rank of the variables in $\sigma$ is $h-1$.
The system is thus acyclic.

The type variables at the left of $\le$ 
are those parameters that come from the result type of a function declaration,
e.g.~$\alpha$ in $nil:list(\alpha)$.
As the result type is a flat type, the variables in a result type are distinct and renamed apart,
hence the variables occurring in a type at the left of $\le$ 
have a unique occurrence in the system. The system is thus trivially left-linear.
\qed\end{proof}

Note that if we allowed contravariant type constructors,
the previous proposition would not hold.

A linear time algorithm for solving acyclic left-linear systems
is given in the next section.

\section{Subtype inequalities}\label{SSI}

The {\em satisfiability of subtype inequalities (SSI) problem} 
is the problem of
determining whether a system of subtype relations\footnote{The SSI problem should not be confused with the semi-unification problem
which is defined with the instantiation pre-ordering, intead of the subtype ordering:
$\exists\Theta\ \bigwedge_{i=1}^n\ \exists \Theta_i\ \tau_i\Theta\Theta_i= \tau'_i\Theta$.
The undecidability of semi-unification is shown in \cite{KTU89}.}
$\bigwedge_{i=1}^n \tau_i\leE \tau'_i$
over types $\tau_1,\tau'_1,...,\tau_n,\tau'_n$
has a solution, i.e., whether
there exists a substitution $\Theta$ such that 
$\bigwedge_{i=1}^n \tau_i\Theta\leE \tau'_i\Theta$.

\begin{definition}
A {\em solution} to an inequality $\tau\leE \tau'$ 
is a substitution $\Theta$ such that $\tau\Theta\leE \tau'\Theta$. 
A {\em maximal solution}
 is a solution $\Theta$
such that for any solution $\Theta'$ there exists a substitution $\rho$
such that $\forall \alpha\in V\ \alpha\Theta'\leE \alpha\Theta\rho$. 
\end{definition}

The SSI problem has been deeply studied in the
functional programming community. Due to the lack of results for the
general case, special instances of the SSI problem have been
identified along several axes: 
\begin{itemize}
\item 
the form of the types: basic types, constructor types, covariant (our case in this paper)
or contravariant;
\item
the structure of the types:
(disjoint union of) lattices \cite{Tiuryn92},
quasi-lattices \cite{Smo89},
n-crown \cite{Tiuryn92},
posets with suprema (our case),
partial orders \cite{Frey97sas};
\item
the form of the type constraints.
\end{itemize}

In this section we show that the type constraints generated by the type checking algorithms
can be solved in linear time in our quite general structure of types,
and that the type constraints generated by
the type inference algorithms can be solved in cubic time,
under the additional assumption that the types form a lattice.

\subsection{The acyclic left-linear case}\label{algo}

We show that the satisfiability of acyclic left-linear subtype inequalities 
can be decided  in linear time,
and admit maximal solutions in our general type structure $(\T,\le)$ of posets with suprema.

In this section, we present an algorithm which proceeds by simplification of 
the subtype inequalities and introduces equations between a parameter and a type.
We say that a system $\Sigma$ is in {\em solved form} if it contains only equations
of the form
$$\{\alpha_1=\tau_1,\ldots,\alpha_n=\tau_n\}$$
where the $\alpha_i$'s are all different and have no other occurrence in $\Sigma$.
The substitution $\Theta_\Sigma=\{\alpha_1\leftarrow \tau_1,\ldots,\alpha_n\leftarrow \tau_n\}$
associated to a system in solved form $\Sigma$
is trivially a maximal solution.
We show that the following simplification rules compute
solved forms for satisfiable acyclic left-linear systems:\\ \\
\begin{tabular}{ll}
(Decomp) & $\Sigma,\ K(\tau_1,...,\tau_m)\le K'(\tau'_1,...,\tau'_n),\ \lra\ \Sigma,\ \bigwedge_{i=1}^n
 \tau_{\iota(i)}\le \tau'_i$\\
& if $K\le K'$ and $\iota=\iota_{K,K'}.$\\
\\
(Triv) & $\Sigma,\ \alpha\le\alpha,\ \lra\ \Sigma$\\
\\
(VarLeft)& $\Sigma,\ \alpha\le \tau,\ \lra\  \alpha=\tau,\ \Sigma[\tau/\alpha]$\\
& if $\tau\not=\alpha$, $\alpha\not\in V(\tau)$.\\
\\
(VarRight) & $\Sigma,\ \tau\le \alpha,\ \lra\  \alpha=\Max(\tau),\ \Sigma[\Max(\tau)/\alpha]$\\
& if $\tau\not\in V$, $\alpha\not\in V(l)$ for any $l\le r\in \Sigma$,
and $\alpha\not\in V(\Max(\tau))$.\\
\end{tabular}\\

\begin{lemma}\label{linear}
The rules terminate in $O(n)$ steps, where $n$ is the sum of the sizes
of the terms in the left-hand side of inequalities. 
\end{lemma}
\begin{proof}
It suffices to remark that each rule strictly decreases the sum of the size of the terms 
in the left-hand sides of the inequalities:
(Triv) and (VarLeft) by one,
 (Decomp) by at least one, and (VarRight) by the size of $\tau$.
\qed\end{proof}

One can easily check that
each rule preserves the left-linearity as well as the 
acyclicity of the system, moreover:

\begin{lemma}\label{soundness}
Each rule preserves the satisfiability of the system,
as well as its maximal solution if one exists.
\end{lemma}
\begin{proof}
Rules (Decomp) and (Triv) preserve all solutions,
by definition of the subtyping order. 
Rule (VarLeft) replaces a parameter $\alpha$ by its upper bound $\tau$.
As the system is left-linear this computes the maximal solution for $\alpha$,
and thus preserves the maximal solution of the system if one exists. 
Rule (VarRight) replaces a parameter $\alpha$ having no occurrence in the left-hand
side of an inequality, hence having no upper bound,
by the maximum type of its lower bound $\tau$;
this computes the maximal solution for $\alpha$,
and thus preserves also the maximal solution of the system if one exists.
\qed\end{proof}

\begin{theorem}\label{leftlin}
Let $\Sigma$ be an acyclic left-linear system.
Let $\Sigma'$ be a normal form of $\Sigma$.
Then $\Sigma$ is satisfiable iff $\Sigma'$ is in solved form,
in which case $\Theta_{\Sigma'}$ is a maximal solution of $\Sigma$.
\end{theorem}
\begin{proof}
Consider a normal form $\Sigma'$ for $\Sigma$. If $\Sigma'$ contains a non variable pair
$\tau\le\tau'$, as this inequality is irreducible by (Decomp) $\Sigma'$ has no solution, 
hence $\Sigma$ is unsatisfiable by lemma \ref{soundness}.
Similarly $\Sigma'$ has no solution if it contains an inequality
$\alpha\le\tau$ with $\alpha\in \vars(\tau)$ and $\tau\not=\alpha$ (prop.~\ref{corfini})
or an inequality $\tau\le \alpha$ with $\alpha\in \vars(\Max(\tau))$ and $\tau\not=\alpha$ (Prop.~\ref{corfini}).
In the other cases, 
by irreducibility and by acyclicity, $\Sigma'$ contains no inequality,
hence $\Sigma'$ contains only equalities that are in solved form, and the substitution associated to $\Sigma'$
is a maximal solution for $\Sigma$.
\qed\end{proof}


\subsection{The general case}


In absence of subtype relations between type constructors of different arities, 
checking the consistency of general
subtype inequalities in finite types has been shown by Frey \cite{Frey97sas}
Pspace-complete in an arbitrary poset,
with a generalization of Fuh \& Mishra's algorithm \cite{FM88}.

It is an open problem whether 
the technique used by Frey for proving
consistency in arbitrary posets can be generalized
to our case with subtype relations between type constructors
of different arities. 

If we assume however that the subtyping relation is a lattice,
it has been shown by Pottier \cite{Pottier00ic} that the satisfiability of subtype inequalities
can be checked in cubic time in the
structure of infinite regular trees, i.e.~recursive types \cite{AC93}.
Note that recursive types admit solutions to equations of the form $\alpha=list(\alpha)$,
namely the type $list(list(...))$.
Below we present Pottier's algorithm 
by a set of simplification rules,
and show that in acyclic systems
the solving of (covariant) subtype constraints on infinite types 
is equivalent to the solving on finite types.

We assume that the structure of type constructors $(\K,\le)$ is a lattice with $\bot$ and
$\top$ types. We maintain our previous assumption on decreasing arities,
except on $\bot$ which is below all (n-ary) type constructors.
We also assume that if $K''/n=glb(K,K')$ then $\range(\iota_{K'',K})\cup\range(\iota_{K'',K'})=[1,n]$,
that is greatest lower bounds do not introduce new parameters.
Similarly, if $K''/n=lub(K,K')$ then $\range(\iota_{K,K''})\cup\range(\iota_{K',K''})=[1,n]$,
Note that there is no loss of generality with this assumption as
the lattice of type constructors can always be completed 
by introducing glb and lub constructors with the right number of parameters.

We consider systems of subtype inequalities between parameters
of flat types, that is of the form $\alpha\le\beta$, $K(\alpha_1,...\alpha_n)\le\alpha$
or $\alpha\le K(\alpha_1,...\alpha_n)$. Non flat types can be represented in this form
by introducing new parameters and inequalities between these parameters and the type they represent.
The simplification rules are the following:

\begin{tabular}{ll}
(Trans) & $\Sigma,\ \alpha\le \beta,\ \beta\le \gamma\ \lra\ \Sigma,\ \alpha\le \beta,\ \beta\le \gamma,\ \alpha\le \gamma$\\
&if $\alpha\le \gamma\not\in\Sigma$ and $\alpha\neq\gamma$.\\
\\
(Clash) & $\Sigma,\  K(\alpha_1,...,\alpha_m)\le \alpha,\ \alpha\le  K'(\alpha'_1,...,\alpha'_n)\ \lra\ \false$\\
& if $K\not\le K'$.\\
\\
(Dec) & $\Sigma,\  K(\alpha_1,...,\alpha_m)\le \alpha,\ \alpha\le  K'(\alpha'_1,...,\alpha'_n)\ \lra$\\
&$\Sigma,\  K(\alpha_1,...,\alpha_m)\le \alpha,\ \alpha\le  K'(\alpha'_1,...,\alpha'_n),\ \bigwedge_{i=1}^n\alpha_{\iota(j)}\le \alpha'_j$\\
\\
& if $K\le K'$, $\iota=\iota_{K,K'}$ and $\{\alpha_{\iota(j)}\le \alpha'_j\}_{j=1,...,n}\not\subset\Sigma$.\\
\\
(Glb) & $\Sigma,\ \alpha\le \beta,\  \alpha\le K(\alpha_1,...,\alpha_m),\ \beta\le  K'(\alpha'_1,...,\alpha'_n)\ \lra$\\
&$\Sigma,\ \alpha\le \beta,\ \alpha\le K''(\alpha''_1,...,\alpha''_l),\ \beta\le  K'(\alpha'_1,...,\alpha'_n), \bigwedge_{j\in J}\alpha''_{\iota'(j)}\le \alpha'_j$\\
\\
& if $K''=\glb(K,K')$ and $K''\not= K$ or $\{\alpha''_{\iota'(j)}\le \alpha'_j\}_{j\in J}\not\subset\Sigma\cup\{\alpha\le \beta\}$\\
&where $\iota=\iota_{K'',K},\ \iota'=\iota_{K'',K'},\ J=\{j|\iota'(j)\not\in\range(\iota)\}$ and\\
& for all $1\le k\le l,\ \alpha''_k=\alpha_i$ if $\iota(i)=k$, $\alpha''_k=\alpha'_j$ if $j\in J$ and $\iota'(j)=k$.\\
\\
(Lub) & $\Sigma,\ \alpha\le \beta,\  K(\alpha_1,...,\alpha_m)\le\alpha,\ K'(\alpha'_1,...,\alpha'_n)\le\beta\ \lra$\\
&$\Sigma,\ \alpha\le \beta,\ K(\alpha_1,...,\alpha_m)\le\alpha,\ K''(\alpha''_1,...,\alpha''_l)\le\beta, \bigwedge_{i\in I}\alpha_{\iota(i)}\le\alpha''_i$\\
\\
& if $K''=\lub(K,K')$ and $K''\not= K'$ or $\{\alpha_{\iota(i)}\le \alpha''_i\}_{i\in I}\not\subset\Sigma\cup\{\alpha\le \beta\}$\\
& where $\iota=\iota_{K,K''},\ \iota'=\iota_{K',K''},\ I=\{i\in[1,m]|\iota(i)\not\in\range(\iota')\}$ and\\
& for all $1\le k\le l,\ \alpha''_k=\alpha'_j$ if $\iota'(j)=k$, $\alpha''_k=\alpha_i$ if $i\in I$ and $\iota(i)=k$\\
\end{tabular}\\

Rule (Trans) computes the transitive closure of inequalities between parameters
and is mainly responsible for the cubic time complexity.
Rule (Clash) checks the consistency of the lower and upper bounds of parameters.
Rule (Dec) decomposes flat types.
Rule (Glb) and (Lub) compute the greatest lower bound of upper bounds of parameters
and the least upper bound of lower bounds.
We remark that if the algorithm is applied to an initial system $\Sigma$
containing a unique inequality of the form $\tau\le \alpha$ and $\alpha\le\tau'$
for each parameter $\alpha$, the algorithm maintains a unique upper and lower bounds
for each parameter. We note $\lb(\alpha)$ (resp. $\ub(\alpha)$)
the lower (resp. upper) bound of $\alpha$ in the system in irreducible form.

\begin{proposition}
The rules terminate.
\end{proposition}
\begin{proof}
Termination with rule (Clash) is trivial. For the other rules,
let us consider, as complexity measure of the system, the couple of integers $(t,e)$ ordered 
by lexicographic ordering, where $e$, the ``entropy'' of the system, is the number $2^v-n$,
where $v$ is the number of parameters in the system, $n$ is the number of inequalities between
parameters, and where $t$, the ``temperature'' of the system, is the sum of the height of constructors
at the right of $\le$, and of the depth of constructors at the left of $\le$. 
The height (resp. depth) of a constructor is its distance to $\bot$ (resp. $\top$) in $(\K,\le)$.
We show that no rule increases the temperature of the system, and each rule
either decreases $t$ or $e$.

Rule (Trans) does not change $t$ and decreases $e$ by 1,
Rule (Dec) does not change $t$ and decreases $e$ by at least 1,
Rules (Glb) either decreases $t$ if $K''\not=K'$ or decreases $e$ otherwise,
and similarly for rule (Lub).
Hence the algorithm terminates.
\end{proof}

\begin{theorem}\cite{Pottier00ic}
A system of inequalities is satisfiable over infinite regular trees
if and only if the simplification rules do not generate $\false$,
in which case the identification of all parameters to their upper bound $\ub(\alpha)$
(resp.~their lower bound $\lb(\alpha)$)
provides a maximum (resp. minimum) solution.
\end{theorem}

Furthermore, one can show that
in our setting of acyclic systems and covariant constructor types,
the solving of subtype constraints on infinite types 
is equivalent to the solving on finite types.

\begin{theorem}
An acyclic system of inequalities is satisfiable over finite types
if and only if the simplification rules do not generate $\false$,
in which case the identification of all parameters to their upper bound
(resp.~lower bound)
provides a maximum (resp. minimum) finite solution.
\end{theorem}
\begin{proof}
It is sufficient to remark that 
the simplification rules preserve the acyclicity
of the system, and that in an acyclic system,
the identification of the parameters to their bounds creates
finite solutions.
\end{proof}

\begin{corollary}
In a lattice structure without $\bot$,
an acyclic system of inequalities is satisfiable over finite types
if and only if the simplification rules do not generate $\false$
and $\ub(\alpha)\not=\bot$ for all parameters $\alpha$.
\end{corollary}

\section{Type inference}\label{TI}

As usual with a prescriptive type system,       
type reconstruction algorithms can be used to omit type declarations
in programs, and still check the typability of the program
by the possibility or not to infer the omitted types \cite{LR91} .
Below we describe algorithms for inferring the type of variables and
predicates, assuming type declarations for function symbols.

\subsection{Type inference for variables}

Types for variables in CLP clauses and queries can be inferred by introducing unknowns
for their type in the variable typing, and by collecting the subtype
inequalities along the derivation of the type judgement just like in the type checking algorithm.

It is easy to check that the system of subtype inequalities thus collected
is still acyclic, as the unknown types for CLP variables appear only in left positions.
The system is however not left-linear if a CLP variable has more than
one occurrence in a clause or a query.

The second algorithm of the previous section can thus be used
to infer the type of variables in CLP clauses and queries.

\subsection{Type inference for predicates}

Types for predicates can be inferred as well
under the assumption that predicates are used monomorphically inside
their (mutually recursive) definition \cite{LR91}.
This means that inside a group of mutually recursive clauses,
each occurrence (even in the body of a clause) of a predicate defined in these clauses must
be typed with rule {\em Head} instead of rule {\em Atom}.
The reason for this restriction, similar to the one
done for inferring the type of mutually recursive functions in ML,
is to avoid having to solve a semi-unification problem:
i.e.~given a system of types $\tau_i,\ \tau'_i$ for $i,\ 1\le i\le n$,
finding a substitution $\Theta$ such that for all $i$
there exists a substitution $\Theta_i$ s.t.~$\tau_i\Theta\Theta_i=\tau'_i\Theta$,
that is proved undecidable in \cite{KTU89}.

Note that the SSI obtained by collecting the subtype inequalities
in the derivation of typing judgements
is still acyclic, as the unknown types for predicates appear only in the right-hand
sides of the inequalities. The second algorithm of the previous section can thus be used
also to infer the type of predicates in CLP programs
under the assumption that the structure of types is a lattice without $\bot$.

One consequence of the acyclicity of the system however, is that the
maximum type of a predicate is always $\top$.
Indeed in our type system a predicate can always be typed as maximally permissive.
In the more general structure of posets with suprema,
unless the unknown types for predicates are compared
with types belonging to different $\le$-connected components (in which case the
predicate is not typable), the substitution of an unknown type by the root of its $\le$-connected component
is always a solution. 
But in all cases, this is obviously not a very informative type to infer.

Our strategy is to infer two types for predicates: the minimum type of the predicate
and a heuristic type. The type inference algorithm proceeds as follows:

\begin{itemize}
\item
Firstly, the minimum type of the predicate is obtained by computing the minimum solution of the SSI
associated to the typing of the complete definition of the predicate.
The minimum type of the $i$th argument of the predicate is the type $\underline\tau_i=\lb(\alpha_i)$
where $\alpha_i$ is the unknown type associated to the $i$th argument of the predicate in the SSI.
This minimum type is a lower bound of all possible typings of the predicate.
\item
Secondly, the heuristic type is computed. This type can be parametric.
It is computed in two steps:
\begin{itemize}
\item
First a heuristic upper type is computed for the predicate.
The heuristic upper type $\overline\tau_i$ of the $i$th argument of the predicate is obtained by collecting
the upper types $\{\ub(\tau_{X_1}),...,\ub(\tau_{X_n})\}$ of
all the variables $\{X_1,...,X_n\}$ which occur in the $i$th position of the predicate in 
its definining clauses.

Let $\tau=\glb\{\ub(\tau_{X_1}),...,\ub(\tau_{X_n})$ be the greatest lower bound of
the types of the variable arguments. We set 

\begin{tabular}{ll}
$\overline\tau_i$&$=\ \top$ if $\tau=\top$ and $\underline\tau_i=\bot$,\\
&$=\ \underline\tau_i$ if $\tau=\top$ and $\underline\tau_i\not=\bot$,\\
&$=\ \top$ if $\underline\tau_i\not\le\tau$,\\
&$=\ \top$ if the identification $\tau_i=\tau$ creates a cycle,\\
&$=\ \tau$ otherwise.\\
\end{tabular}

\item
Then the heuristic type is computed by inferring a possibly parametric type from the
SSI associated to the heuristic upper type.
The candidates for parametric types are the parameters bounded by $\bot$ and $\top$
in the SSI associated to the heuristic upper type.
Each candidate is checked iteratively by replacing it with a new constant and by identifying
all parameters which have the new constant in one of their bounds.
\end{itemize} 
\end{itemize} 

Although tedious, one can easily check that the conditions imposed 
in the definition of the heuristic type create sound typings.
The heuristic types thus provide correct type declarations for type checking the program.

\section{Implementation of the type system}\label{implementation}

\subsection{The Wallace library for solving subtype inequalities}

Our current implementation uses the Wallace library by F.Pottier \cite{Pottier00wallace} 
for solving the subtype inequalities for type inference and type checking.
In both cases, the set of type constructors $(\K,\le)$ has thus to be a lattice
as described in section \ref{SSI}.
Note that the type system did not require that condition: $\leq$
could be any arity decreasing order relation on $\K$.

As required in the type inference algorithm,
the $\top$ element is distinguished from the type {\em term} which stands for all Prolog terms.
The type $\bot$ is not considered as a valid typing
as it is an empty type.

Note that the Wallace library authorizes constrained type
schemes, like for example $+:\forall \alpha\leq \float\ \alpha\times\alpha\ra\alpha$,
which expresses the resulting type of $+$ as a function of the type of its arguments.
For the sake of simplicity, we do not consider constrained type schemes in this paper.

\subsection{The type checker}

The type checker first reads the Prolog files and deduces the files containing 
type information to load. There is one file for each Prolog file source plus 
one file for each module used (as \verb|:- use_module(somemodule)| in 
Sicstus Prolog). The system then loads the type files and builds the
structure of type constructors. 

The type checker does not impose to give the type of CLP variables in clauses and queries.
Instead the type of variables is inferred as described in section \ref{TI}.
The environment $U$ is built with type unknowns for variables. 
The subtype inequality system is collected by applying the rules of the type system
and at each step, Wallace is used to solve the type constraints.

One difficulty appears for checking the definitional genericity condition.
A type error must be raised when the definition of a predicate uses, as argument 
of the head of the clause, a term whose type $\tau$ is a subtype of an 
{\em instance} of the declared type $\tau '$ for this argument, and not just of a 
{\em renaming}. But Wallace is not able to make the difference between being 
a subtype of an instance or of a renaming of a type $\tau'$. The following 
consideration allows us to work around this difficulty. If $\tau$ is a 
subtype of a renaming of $\tau'$, for all instances $\tau'\Theta'$ of $\tau'$ 
there must exist an instance $\tau\Theta$ of $\tau$ such that $\tau\Theta\leq\tau'\Theta'$. 
For checking definitional genericity,
we thus replace each parameter $\alpha$ appearing in the declared type of the head predicate
by a constructor $\kappa_\alpha$, that does not appear in the program, and
such that~:
\begin{itemize}
\item $\kappa_\alpha\leq term$, and $\kappa_\alpha\not\leq \mu$ 
        for all constructor $\mu$, $\mu\neq\kappa_\alpha$, $\mu\neq term$
\item  $\mu\not\leq\kappa_\alpha$ 
        for all $\mu\neq\kappa_\alpha$.
\end{itemize}

If the rule {\em (Atom)} can be applied, using the transformed type, then the 
rule {\em (Head)} can be applied as well with the original type.


\subsection{Type inference for predicates}\label{sec:type-inf}

As described in section \ref{TI}, two types are infered for predicates:
a minimum type which is a lower bound of all possible typings of the predicate,
and a heuristic type which may be parametric.

If type inference is just displayed for user information, we print both types. 
If it is used for typing automatically the program in a non-interactive manner, then we choose the heuristic 
bound, since it is the most permissive type.


\section{Experimental results}\label{boum}
\subsection{Detection of programming errors}
Here we show a small catalog of the kind of programming errors detected by the type checker.

\subsubsection{Inversion of arguments in a predicate or a function}
This error can be detected when, for example, a variable occurs in two positions 
that have incompatible types. 
\begin{example}
Consider the following clause where the arguments of the length predicate have been 
reversed.
\begin{center}
\tt p(L1,L2,N) :- append(L1,L2,L3),length(N,L3).
\end{center}
with the usual declarations~:
\begin{list}{}{}
\item {\tt append : }$list(\alpha)\times list(\alpha)\times 
      list(\alpha)\rightarrow \pred$
\item {\tt length : }$list(\alpha)\times int \rightarrow \pred$
\end{list}
By the rule {\em (Atom')}, the variable {\tt L3} must be of both types $list(\alpha)$ and $int$.
In the type hierarchy we use, there is no type smaller than $list$ and $int$.
The subtype inequalities in the premise of rule {\em (Atom')} are thus unsatisfiable and
a type error is raised.
\end{example}

Note that this example motivates the discard of type $\bot$~: 
otherwise, no error would be detected on variables, 
since the empty type $\bot$ could always be inferred for the type of any variable.

\subsubsection{Misuse of a predicate or a function}
This error is detected when a term of a type $\tau$ appears as an argument 
of a predicate, or of a functor that expects an argument of type $\tau'$, 
but $\tau\not\leq\tau'\Theta$ for any substitution $\Theta$.
\begin{example}
Consider the following clause~:
\begin{center}
\tt p(X,Y) :- Y is (3.5 // X).
\end{center}
With type declarations:
\begin{list}{}{}
\item{\tt '//' : } $int \times int \rightarrow int$ for integer division,
\item{\tt is : } $float \times float \rightarrow \pred$.
\end{list}
We try to use a {\em float} ({\tt 3.5}) where an {\em int} is expected. 
The rule {\em (Atom')} does not apply.
\end{example}

This kind of error can be detected also inside call to foreign predicates,
through the Prolog interface with the C programming language for example.
\begin{example}
Consider the declaration of a predicate {\tt p} defined in C using the
Sicstus - C interface~:
\begin{list}{}{}
\item {\tt foreign(p, p(+integer)).}
\end{list}
Such a declaration is interpreted as a type declaration for {\tt p}~:
\begin{list}{}{}
\item {\tt p : }$int \rightarrow \pred$
\end{list}
Then a call in a program such as~:
\begin{list}{}{}
\item {\tt :- p(3.14).}
\end{list}
raises a type error since the argument is a $float$ and the predicate
expects an $int$.
\end{example}

\subsubsection{Wrong predicate definition w.r.t. the declared type}
This error is detected by two ways, corresponding to the two preceding kinds of errors.
In the two following examples, the predicate {\tt p} has been declared with
type $int \rightarrow \pred$.
\begin{example} Let {\tt p} be defined by 
{\tt p([]).}\\
Here the term {\tt []} is used as an argument of {\tt p}, which requires that 
{\tt p} accepts arguments of type $atomic\_list$. But 
$atomic\_list\not\leq int$ and the rule {\em (Atom')} does not apply.
\end{example}
\begin{example}
Let {\tt p} be defined by~:
\begin{list}{}{}
\item {\tt p(X) :- length(X,2).}
\item with {\tt length : }$list(\alpha) \times int \rightarrow \pred$
\end{list}
In this case, we will infer a type for {\tt X} that must be smaller than 
$list(\alpha)$ (using rule {\em (Atom')}, because {\tt X} is used by 
{\tt length}) and smaller than $int$ (using the rule {\em (Head')}). 
As before, these types have no common subtypes, and an error is raised.
\end{example}

\subsubsection{Violation of the definitional genericity condition}

\begin{example}
Let~:
\begin{list}{}{}
\item {\tt  p([1]).}
\item with {\tt p : }$list(\alpha)\rightarrow \pred$
\end{list}
Although the argument of {\tt p} is a list, but its type is $list(int)$, an
{\em instance} and not a {\em renaming} of $list(\alpha)$ (because $int\not\leq\kappa_\alpha$). 
\end{example}
This error can also be detected when a variable is in the head of the clause~:
\begin{example}
Let~:
\begin{list}{}{}
\item {\tt p([X]) :- X < 1.}
\item with~:
  \begin{list}{}{}
  \item {\tt p : }$list(\alpha)\rightarrow \pred$
  \item {\tt < : }$float \times float \rightarrow float$
  \end{list}
\end{list}
The variable {\tt X} must be of type $float$ and $\kappa_\alpha$. The only 
common subtype is $\bot$ and an error is raised.
\end{example}
\subsection{Type checked programs}
To test our system, we first tried it on 20 libraries of Sicstus Prolog, that is around 600 predicates. 
Then we type checked an implementation of CLP(FD) written completely in Prolog, using a lot of meta-predicates,
that contains around 170 predicates. 
These tests where done using type declarations for around 100 built-in ISO Prolog predicates and 
for some more built-in Sicstus predicates.

Some type errors obtained in the libraries came from the overloading of some function symbols.
For example,
the function {\tt '-'/2} is used for coding pairs as well as for coding the arithmetic operation over numbers. 
Another example of overloading comes from options~: it happens that some terms are common to two sets of options,  
of types $\tau_1$, $\tau_2$. 
In this case, it is enough to create a subtype $\tau$ of both $\tau_1$ and $\tau_2$, 
and tell that the common terms are of type $\tau$.

We also skip the type checking of some particular declarations, such as mode declarations
(which are not used by our type system)~:
\begin{example}\ 
\begin{list}{}{}
  \item {\tt :- mode p(+,-,+) , q(-,?).}
\end{list}
\end{example}
These declarations can be typed in another type structure for mode declarations, 
but not in the same type structure as the one for predicates, since the predicate symbols {\tt p}, {\tt q},
{\tt +}, {\tt -}
are clearly overloaded in such declarations.

\subsection{Type inference for predicates}

As said in section (\ref{sec:type-inf}), we infer an interval of types
for predicates. 
Both bounds of the interval may offer interesting information.

\begin{example}
\begin{verbatim}
append([Head| Tail], List, [Head| Rest]) :- 
    append(Tail, List, Rest).
append([], List, List).

Minimum type: list(bottom), list(bottom), list(bottom) -> pred
Heuristic infered type:  list(A), list(A), list(A) -> pred
\end{verbatim}
\end{example}

\begin{example}
\begin{verbatim}
sum_list([], Sum, Sum).
sum_list([Head| Tail], Sum0, Sum) :- 
    Sum1 is Head+Sum0, sum_list(Tail, Sum1, Sum).

Minimum type: list(bottom), bottom, bottom -> pred
Heuristic infered type: list(float), float, float -> pred
\end{verbatim}
\end{example}

Sometimes, the heuristic infers a too permissive type.
This is in particular the case with overloaded arithmetic predicates expressions,
that are always typed as {\tt float}, not {\tt int}.

\begin{example}
\begin{verbatim}
length([],0).
length([_|Tail],R) :- length(Tail, L), R is L+1.

Minimum type:  list(bottom), int -> pred
Heuristic infered type:  list(A), float -> pred
\end{verbatim}
\end{example}

The heuristic may also infer a type which is too restrictive.

\begin{example}
\begin{verbatim}
is_list(X) :- var(X), !, fail.
is_list([]).
is_list([_|Tail]) :- is_list(Tail).

Minimum type:  list(bottom) -> pred
Heuristic infered type:  list(A) -> pred
\end{verbatim}
This is a typical example where the maximum type, here 
\begin{verbatim}
is_list: term -> pred
\end{verbatim}
is in fact the intended type.
\end{example}

These examples should clearly justify the heuristic approach to type inference
for predicates in a prescriptive type system.

Finally, the interesting $flatten$ predicate illustrates the
remarkable flexibility of the type system.

\begin{example}
\begin{verbatim}
flatten([],[]) :- !.
flatten([X|L],R) :- !, flatten(X,FX), flatten(L,FL), append(FX,FL,R).
flatten(X,R) :- R=[X].

Minimum type : list(bottom), list(bottom) -> pred
Heuristic infered type : term, list(term) -> pred
\end{verbatim}
\end{example}


\subsection{Benchmarks}

The following table sums up our evaluation results.
The first column indicates the type checked Prolog program files.
The second column indicates the number of predicates defined in each file first, and then the maximum number of atoms by clause and by complete connected component.
The third column indicates the CPU time in seconds for type checking the program
with the type declarations for function and predicate symbols.
The fourth column indicates the CPU for inferring the types of predicates
with the type declarations for function symbols only.
The last column indicates the percentage of predicates for which
the infered type is exactly the intended type.

The last test file is another implementation of CLP(FD) on top of prolog
which uses a lot of metaprogramming predicates.

\begin{table}
\begin{center}
\begin{tabular}{|l|c|c|c|c|c|}
\hline
File       &  \# predicates & Type Checking & Type Inference &  \% exact types  \\
& max \# atoms&&&\\
\hline
arrays.pl  & \ 13\ \ \ \ 9/16\ &   2.18 s &   11.91 s & 23 \%\\
assoc.pl   & \ 31\ \ \ 11/24\ &   5.29 s &   40.13 s & 68 \%\\
atts.pl    & \ 14\ \ \ 20/119 &   7.43 s &   77.47 s & n/a \\
bdb.pl     & 101\ \ \ 27/27\ &  23.56 s &   41.10 s & 64 \%\\
charsio.pl & \  15\ \ \ \ 7/7\ \ &   1.27 s &    2.21 s & 33 \%\\
clpb.pl    & \  59\ \ \ 20/77\ &  24.35 s & 1827.32 s & n/a \\
clpq       & 396\ \ \ 39/160 & 355.12 s & 4034.37 s & n/a \\    
clpr       & 439\ \ \ 39/160 & 304.45 s & 3958.41 s & n/a \\
fastrw.pl  & \ \   4\ \ \ \ 5/7\ \ &   0.44 s &    0.76 s & 100 \%\\
heaps.pl   & \  21\ \ \ \ 8/18\ &   3.49 s &   43.33 s & 71 \%\\
jasper.pl  & \  32\ \ \ 11/11\ &   7.43 s &   11.97 s & 84 \%\\
lists.pl   & \  39\ \ \ \ 6/9 \ \ &   2.23 s &   16.17 s & 97 \%\\
ordsets.pl & \  35\ \ \ \ 7/18\ &   7.43 s &  199.38 s & 97 \%\\
queues.pl  & \  12\ \ \ 11/18\ &   1.37 s &    4.12 s & 75 \%\\
random.pl  & \  11\ \ \ 18/18\ &   2.43 s &    4.12 s & 55 \%\\
sockets.pl & \  24\ \ \ 15/27\ &   6.79 s &   15.43 s & n/a \\
terms.pl   & \  13\ \ \ 18/27\ &   6.96 s &  308.69 s & 77 \%\\
trees.pl   & \  13\ \ \ \ 6/15\ &   3.07 s &   12.64 s & 31 \%\\
ugraphs.pl & \  87\ \ \ 12/24\ &  48.21 s &  274.22 s & 67 \%\\
\hline
clp-fd.pl & 163\ \ 20/71\ & 24.35 s & 59.65 s & n/a\\
\hline
\end{tabular}
\end{center}
\caption{Benchmarks.}
\end{table}

The same algorithm is used for solving the systems of subtype inequalities
for type checking and type inference.
The difference between computation times
comes from the handling of complete connected components of definitions
for type inference, whereas for type checking, clauses are type checked
one by one. In particular CLP(R) and CLP(Q) have very large mutually recursive clauses.

In the library for arrays, 
the low percentage of exact matches between the infered type and the intended type
is simply due to the typing of indices by $\float$ instead of $int$.
The errors in the other libraries are also due
to the typing of arithmetic expressions by $\float$,
and sometimes to the use of the equality predicate $=_{\alpha,\alpha}$
which creates a typing by $\term$ for some arguments instead of a more restrictive typing.

In the library CLP(FD), finite domain variables are typed with type $int$.
Similarly in the library CLP(R), variables over the reals are typed
with type $\float$. One consequence is that the type checker then allows
coercions from finite domain variables to real constraint variables.
To make these coercions work in practice one modification in the CLP(R) library was necessary.

\section{Conclusion}

Typing constraint logic programs for checking programming errors statically
while retaining the flexibility required for preserving all the metaprogramming
facilities of logic programming and the usual coercions of constraint programming, 
is the challenge that conducted
the design of the type system presented in this paper.
Our experiments with the libraries of Sicstus Prolog have shown that 
the type system is simple and flexible enough to accept a large variety of constraint logic
programs. The main difficulties are located to conflicts of overloading for some 
predicates or functions. Such ad hoc polymorphism
could be resolved by considering disjunctive formula over types \cite{DGS99acsc}.
Examples have been given also to show that the type system is useful enough for detecting
programming errors such as the inversion of arguments in a predicate,
or the unintended use of a predicate.

The price to pay for this flexibility is that our type system may be regarded
as too permissive. Some intuitively ill-typed queries may be not rejected by the type system.
We have analyzed these defects in terms of the subject reduction properties of the type system.
In particular we have shown that the addition of the typing constraints on variables
to well-typed programs and queries
suffices to state subject reduction w.r.t.~both CSLD resolution and substitution steps,
and has for effect to reject a larger set of clauses and queries
by checking the satisfiability of their constraints with the type constraints at compile-time.

The lattice assumption for the type structure, due to the implementation in Wallace
of  subtype constraints, may be regarded also as too demanding in some cases. 
We have already relaxed that assumption by rejecting the bottom element
from the structure of types. Nevertheless the decidability of subtype constraints under
more general assumptions is an interesting open problem.
In particular, whether the method of Frey \cite{Frey97sas} can
be extended to cover subtype relations between type constructors of different arities,
as required in our approach, is an open question.

Finally, it is worth noting that the results presented here are not 
limited to logic programming languages. They should be relevant
to various constraint programming languages, where the main difficulty is
to type check constraint variables, that express the communication between 
different constraint domains.

\vspace{5mm}
\noindent
{\bf Acknowledgment.}
We would like to acknowledge fruitful discussions
with Fran\c{c}ois Pottier, Didier R\'{e}my,
Jan Smaus and Alexandre Frey on this work.
We are also grateful to the referees for their peer reviews.


\end{document}